# Optical properties and Raman studies of partially edge terminated vertically aligned nanocrystalline MoS$_2$ thin film

Anand P.S. Gaur,[1] Satyaprakash Sahoo,[1,*] Majid Ahmadi,[1] Maxime J-F Guinel,[1,2] Sanjeev K. Gupta,[3] Ravindra Pandey,[3] Sandwip K Dey[4] and Ram S. Katiyar[1,*]

[1] Department of Physics and Institute for Functional Nanomaterials, College of Natural Sciences, University of Puerto Rico, PO Box 70377, San Juan, Puerto Rico 00936-8377, USA
[2] Department of Chemistry, College of Natural Sciences, University of Puerto Rico, PO Box 70377, San Juan, Puerto Rico 00936-8377, USA
[3] Department of Physics, Michigan Technological University, Houghton, Michigan 49931, USA
[4] Materials Science and Engineering Program, Arizona State University, Tempe, AZ 85287 USA

**Abstract:** The optical and vibrational properties of nanocrystalline thin films of MoS$_2$, comprised of a mixture of edge terminated vertically aligned (ETVA) and (001)-oriented regions, on large insulating substrates are reported. From high resolution transmission electron microscopy (HRTEM), the average size of ETVA nanocrystals were ~5 nm and each nanocrystal consisted of only 3 to 5 monolayers of MoS$_2$. The films were highly transparent (~80%) but the percent of transmittance decreased as the energy of the incident light approached to the band gap. Additionally, weak excitonic peaks were observed both in the absorption and transmission spectra. The room temperature Raman study showed that both the $E^1_{2g}$ and $A_{1g}$ modes were significantly broader, and a few additional Raman modes were observed when compared to bulk MoS$_2$. The broadening of the $A_{1g}$ mode was analyzed using the phonon-confinement model and the calculated particle size was in good agreement with TEM observations. Moreover, the temperature coefficient of the $A_{1g}$ mode was estimated from the temperature dependent Raman studies.

**\*Corresponding authors:**

Email: rkatiyar@hpcf.upr.edu (R. S. Katiyar), satya504@gmail.com (S. Sahoo)

**Introduction**:

Two-dimensional (2D) materials are appealing for use in next-generation nanoelectronic devices, and to date, graphene has been the most widely studied because of to its wealth of science, superior carrier mobilities and potential for transistor applications.[1, 2] However, due to the lack of bandgap of pristine graphene, the fabrication of its engineered counterpart (i.e., graphene nanoribbons) has resulted in process complexity,[3, 4] reduced mobility,[5, 6] and requirement of high voltages.[7, 8] Therefore, a few layers or a single layer (i.e., a unit cell) of 2D transition metal di-chalcogenides or TMDs (i.e., $MX_2$ where transition metal, M= Mo, W and chalcogen, X= S, Se or Te) could represent the ultimate limit of miniaturization by serving as the transistor channel material for low-power nanoelectronic devices. The prototypical TMD of molybdenum disulfide ($MoS_2$) is of particular interest due to its demonstrated thickness-dependent band gap, and thermal and optical properties; for example, the bulk has an indirect gap of 1.2 eV whereas a monolayer exhibits a direct gap of 1.8 eV).[9-11] Since the dimensions of $MoS_2$ can be less than 1 nm, $MoS_2$-based transistors could lead to smaller and more power-efficient transistors with reduced short channel effects.[12, 13] Additionally, optoelectronic and energy-harvesting applications that require ultra-thin but transparent semiconductors, $MoS_2$ could complement graphene in hybrid structures.[14-16]

To date, most studies on $MoS_2$ have utilized free-standing or substrate-integrated layers with out-of-plane c axis orientation of [001]. In one study on a few layers of $MoS_2$, deposited by a vapor-phase method on $SiO_2$/Si substrate, the in-plane $E_{2g}$ and out-of-plane $A_{1g}$ Raman modes were determined and the phonon contribution to the thermal conductivity was estimated.[17] In contrast, studies on edge terminated vertically aligned (ETVA) structures, i.e., in which (001) planes are perpendicular to the substrate, are rare.[18] Since the top surface or plane of an ETVA structure has dangling bonds, it may be an active plane for catalytic reactions such as oxygen reduction and photo-oxidation of water.[19, 20] Thus, nanocrystalline ETVA of $MoS_2$ could be of high environmental interest since organic chemicals in air may be eliminated via the utilization of solar radiation. Therefore, for electronic or catalytic applications, the fundamental optical and vibrational properties of such nanocrystalline materials are pertinent.

In this report, the optical and temperature-dependent Raman scattering properties of nanocrystalline $MoS_2$ layers, exhibiting a mixture of ETVA and (001)-oriented regions are discussed. The $MoS_2$ layers were synthesized by the rapid sulfurization of Mo metal films on insulating substrates, and the $A_{1g}$ Raman mode was analyzed using a phonon confinement model, from which the estimated grain size was compared with transmission electron microscopy (TEM) observations. Additionally, first principles calculations were carried out to understand the absorption and phonon dispersion characteristics for bulk and single-layered $MoS_2$.

**Results and Discussions:**

Samples were grown by the hydro-sulfurization of molybdenum-coated $SiO_2$/Si and double side polished $Al_2O_3$ substrates at 550 ºC for 30 minutes in vacuum. Elemental sulfur powder was placed in a horizontal tube furnace under atmospheric pressure with a (Ar+$H_2$) gas mixture. Raman measurements were carried out using a Horiba-Jobin T64000 micro-Raman system, and a 532 nm wavelength excitation radiation from a diode laser. The $MoS_2$ samples were transferred to conventional copper TEM grids, and characterized using an analytical transmission electron microscope (TEM: JEOL JEM-2200FS) optimally configured for energy filtered imagery, chemical analysis, and diffraction pattern techniques.

Theoretical calculations were performed within the density functional theory formalism to study bulk and monolayer of $MoS_2$. Here, the Quantum Espresso code employed the local density approximation (LDA), via the adoption of the exchange-correlation function of Perdew and Wang's, to density functional theory.[21,22] An ultrasoft pseudopotential description of the electron-electron interaction was used with valence electrons $4d^5$, $5s^1$ and $3s^2$, $3p^4$ of Mo and S atoms, respectively. A plane-wave basis set for the electronic wave functions and the charge density was applied, with kinetic energy cutoffs set to 50 Ry and 500 Ry, respectively. For the monolayer, the periodic boundary condition of the separation between neighboring cells in (xy) plane was a distance of 20 Å. The configurations using the criteria of forces and stresses on atoms were relaxed until the energy change was less than $10^{-4}$ eV. For the geometry optimization, all the internal coordinates were relaxed until the Hellmann-Feynman forces were less than 0.01 eV/Å. Additionally, phonon calculations for the $MoS_2$ bulk were performed by density functional perturbation theory (DFPT) with fixed occupation scheme for electronic excitation.[23] A 11×11×1 Monkhorst-Pack (MP) k-mesh was found to yield phonon frequencies converged to within 4 $cm^{-1}$ and 5×5×1 q-mesh in the first BZ was used in the interpolation of the force constants for the phonon dispersion calculations. The convergence of phonon calculations was also tested for higher mesh and cutoff energy and termed that the phonon frequencies converged up to 4 $cm^{-1}$. The phonon density of states were calculated using a 11×11×1 k-point mesh, and was found to yield convergent results.

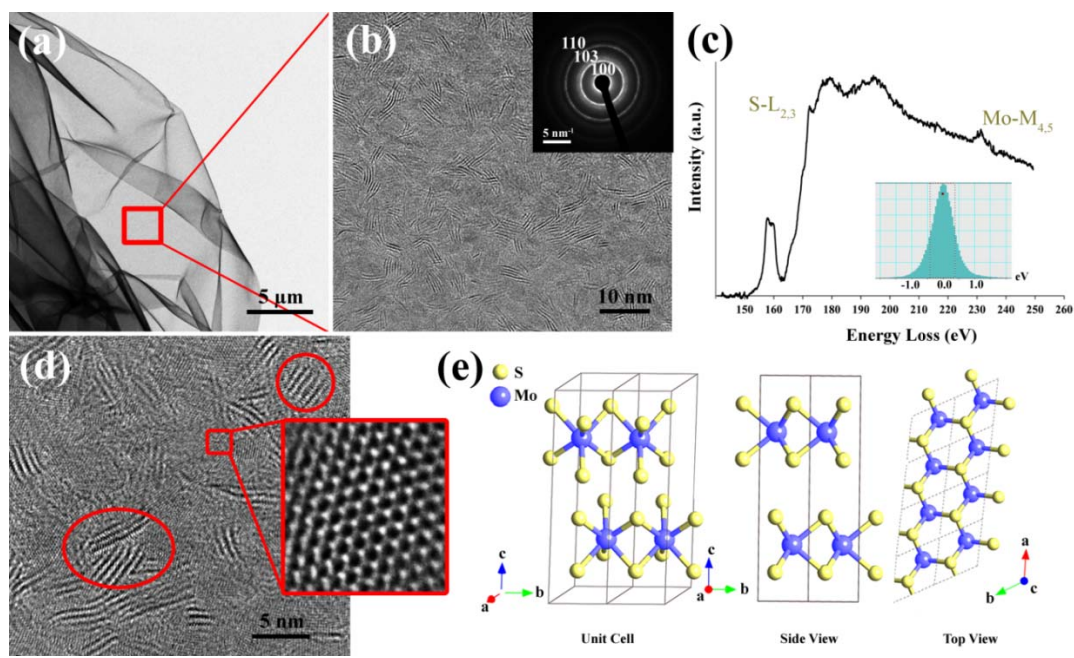

**Figure 1. (a) Bright-field TEM image of suspended MoS$_2$ film, (b) HRTEM image of MoS$_2$, (c) EELS spectrum with core-loss and fine structure for Mo and S edges (resolution ~ 1 eV), and (d) HRTEM image of the ETVA structure, and for comparison, (e) schematics of 2H MoS$_2$ lattice structure. The insets in b, c, and d shows corresponding SAED pattern, zero loss peak, and enlarged HRTEM image, respectively.**

Figure **1** (a) is a bright-field TEM image of a large, free-standing MoS$_2$ and Fig. 1 (b) shows an high-resolution TEM (HRTEM) image of the same film with a typical selected area electron diffraction (SAED) pattern in the inset, indexed to hexagonal MoS$_2$ (JCPDS card No. 73-1508). The unique structure of the film is comprised of dispersion of ETVA-structured and (001)-oriented regions of MoS$_2$ nanocrystals, with densities of the two variants (equally focused at the same depth) in similar proportions. Note, the overall film is flat and ETVA nanocrystals do not protrude from the film, with average thickness and lateral width of the ETVA regions of about 5 nm and 6 nm, respectively. A HRTEM image showing the regions of the ETVA

nanocrystals is illustrated in Fig. 1(c) with an enlarged image in the inset. The electron energy loss spectra (EELS) were recorded from the same film using the in-column energy filter fitted on the 2200FS. The EELS data in Fig 1 (c), having a resolution of 1 eV, show the S $L_{2,3}$ and Mo $M_{4,5}$ edges located at 165 eV and 227 eV, respectively. The energy calibration of the core loss spectrum was ascertained by aligning the zero loss (inset of Fig. 1 (c)) prior to acquiring any data. An HRTEM image showing clearly the structure of ETVA nanocrystals is displayed in Fig. 1(d). It compares well with the structure of 2H $MoS_2$ for which schematics of the unit cell and views along two directions are displayed in 1 (e).

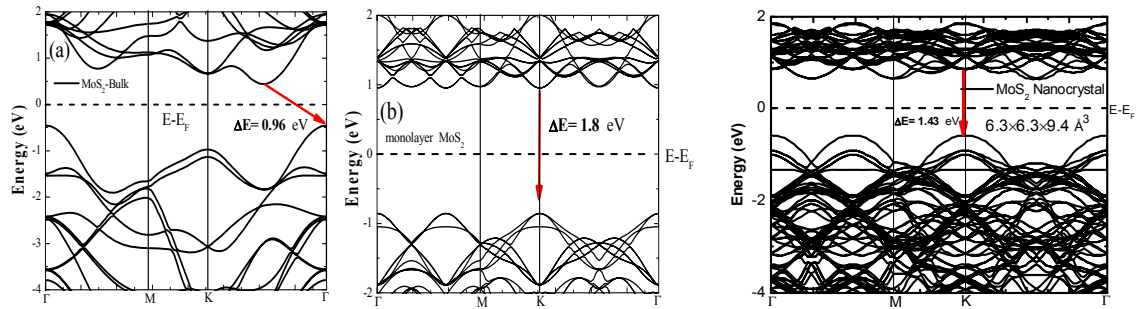

**Fig.2. Calculated band structure of (a) bulk (b) monolayer (c) nanocrystalline $MoS_2$**

The calculated electronic band structures at the level of LDA-DFT approximation along the high symmetry directions in the Brillouin zone are shown in Figs 2 (a), (b) and (c) for bulk, monolayer and nanocrystal of $MoS_2$, respectively. For bulk $MoS_2$, an indirect band gap (1.2 eV), whereas for $MoS_2$ monolayer, a direct band gap (1.8 eV) have been observed; the direct excitonic transition energy at the Brillouin zone K point changes with layer thickness, and becomes so high in a monolayer that the material changes into a 2D direct bandgap semiconductor. Here, the band structure of $MoS_2$ nanocrystals[24] of dimension 9.4 Å, also shows a direct band gap but with a magnitude of ~1.43 eV. It is shown that the direct excitonic transition energy at the Brillouin zone K point changes with layer thickness. Remarkably, the direct transition energy becomes so high in mono-layer MoS$_2$ that the material changes into a two-dimensional direct bandgap semiconductor. The calculated band gap is in good agreement with reported experimental result.[9]

In order to understand the optical properties of ETVA MoS$_2$, the optical absorption and transmission properties (Fig. 3 (a)) of a film grown on a transparent Al$_2$O$_3$ substrate were determined. The transmittance decreases with decrease in wavelength and vice versa for the absorption.

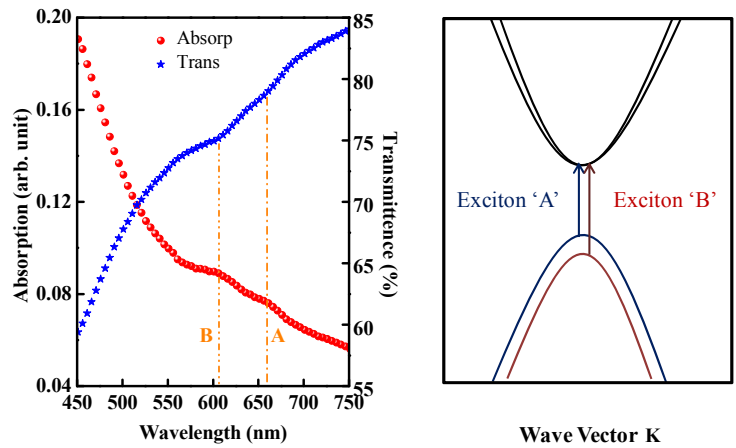

**Fig. 3. (a) Absorption and transmission spectra of a ETVA MoS$_2$ thin film. (b) Schematic representation of A and B exciton transitions.**

The film is quite transparent for light of energies than the band gap of the MoS$_2$. There are two peaks in both the absorption and transmission spectra. These peaks are indicated by two dash-dot lines (A and B) on the plot. These are reported to be the excitonic transitions at the Brillouin zone K point in bulk MoS$_2$ (Figure 3b) appearing at 610 and 675 nm and denoted as A and B transition, respectively.[25, 26] The energy difference between these two transitions is about 0.12 eV and this is due to the spin-orbital splitting of the valence band.[10, 27] Note that in case of bulk MoS$_2$ these peaks are quite strong and sharp whereas for thin films they are considerably weaker and also broader. The broadness in the peaks could be due to the semi crystalline nature of the film. This is in agreement with the diffraction patterns recorded in the TEM. Although the current DFT calculations show that direct excitonic transitions remain unchanged with decreasing number of layers, the band gap, however, shifts from indirect to direct in the case of single layer $MoS_2$ (Fig 2 (b)).

In the following paragraphs, focus is on the phonon properties of $MoS_2$ bulk. Figure 4 presents the phonon dispersion curves along with the phonon density of states for $MoS_2$ bulk. There are eighteen (three acoustic and twelve optical) phonon branches as $MoS_2$ has six atoms per unit cell. Those that vibrate in-plane (longitudinal acoustic, LA, and transverse acoustic, TA) have a linear dispersion and higher energy than the out-of-plane acoustic (ZA) mode. The latter displays a $q^2$-dependence analogous to that of the ZA mode in graphene (which is a consequence of the point-group symmetry)[28] The overall agreement between theory and experiment[29] is good, even for the inter-layer modes. This confirms our expectation that the LDA-DFT approximation describes reasonably well the inter-layer interaction. The low frequency optical modes are found near 50-60 $cm^{-1}$ and correspond to rigid-layer shear/vertical motion, respectively (in analogy with the low frequency optical modes in graphite).[30] When the wave number $q$ increases, the acoustic and low frequency optical branches almost match. The phonon dispersion relations along with phonon density of states possess two distinct regions. The highest vibrational modes are due to S atoms whereas the vibrations of the Mo atoms play the most prominent role for lowest modes. There is a large difference between the frequencies of highest and lowest level vibration modes due to large mass ratio between Mo and S atoms. The high frequency optical modes are separated from the low frequency modes by a gap of ~50 $cm^{-1}$. We have drawn the atomic displacements of the Raman active modes ($E_{2g}^1$ and $A_{1g}$) and the infrared active mode $E_{1u}$. The Raman active modes are also indicated in the phonon dispersion of Fig. 3.

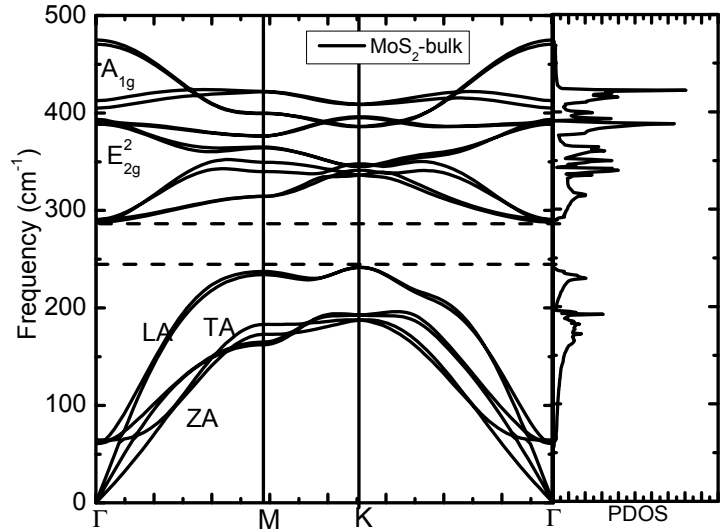

**Fig. 4. Calculated phonon dispersion and the phonon density of states of bulk MoS$_2$.**

For layered materials, Raman spectroscopy has been employed very efficiently to determine the number of layers (using the band position and intensity), the mechanical properties and the thermal properties.[17, 31] Moreover, it is possible to determine the grain size in sub nanometer range decisively.[32] In the following section we present comparative Raman studies of bulk and ETVA MoS$_2$ nanocrystalline thin films at room temperature. Hexagonal MoS$_2$ belongs to space group $D_{6h}$ ($P6_3/mmc$) for which zone centre Raman and Infra red (IR) active vibrational modes are presented in the following irreducible decomposition[33]: $\Gamma = A_{1g} + E_{1g} + 2E_{2g}^1 + 2E_{2g}^2 + E_{1u} + A_{2u}$ where $A_{1g}$ (409 cm$^{-1}$ out of plane breathing mode), $E_{2g}^2$ (32 cm$^{-1}$), $E_{1g}$ (519 cm$^{-1}$ in plane vibration) and $E_{2g}^1$ (383 cm$^{-1}$ interlayer vibration) modes are Raman active and represents out of plane, in plane, and interlayer vibration while rest of two are IR active. Interestingly, $E_{1u}$ and $E_{2g}^1$ modes are degenerate (provided van der Waal interaction between the layers is very weak) and the only difference is that they vibrate in 180º interlayer phase shift[34]. Figure 5 shows the room temperature first order Raman spectra of bulk and nanocrystalline thin films of $MoS_2$.

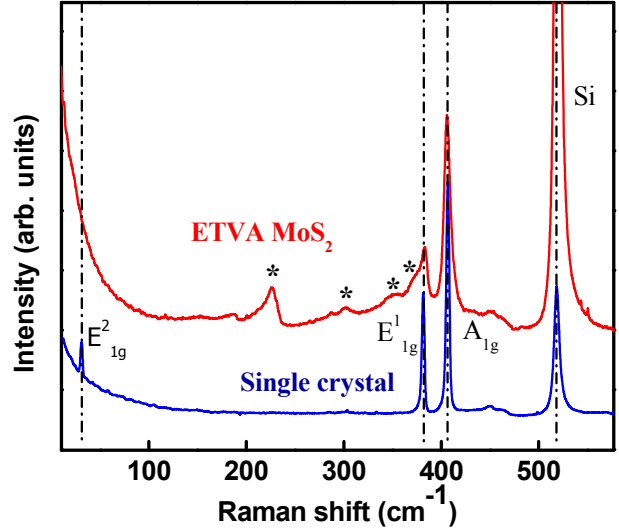

**Fig. 5. Comparison of the room temperature Raman spectra of mechanically exploited bulk $MoS_2$ and ETVA $MoS_2$ thin films. The Si substrate peak is at 521 cm$^{-1}$.**

Sharp Raman bands were observed at 31 cm$^{-1}$, 383 cm$^{-1}$ and 407 cm$^{-1}$ and assigned to $E_{2g}^2$, $E_{2g}^1$ and $A_{1g}$ modes for single crystal while these bands showed broadening in nanocrystalline $MoS_2$. Secondly, there were few weak additional peaks at about 225, 300 and 375 cm$^{-1}$ for ETVA $MoS_2$ films (marked as * in Fig.5). The origin of these peaks is associated with the zone boundary phonons. The long range lattice periodicity is broken grain boundaries causing the relaxation of the Raman selection rule. Consequently, the some zone boundary phonons appeared in the Raman spectra of nanomaterials. The peak positions match with some of the zone boundary (*M* or *K*-symmetry points) of the phonon dispersion curve shown in Fig 4. Similar results were reported for PLD grown $MoS_2$ films at room temperature and they found they disorder peak disappears with annealing of sample[35]. Similar results were reported by Yang *et al.* where the low frequency band was assigned to octahedral MoS$_2$ which is considered metastable decaying[36]. An interesting feature lies in the comparison of the intensity of the substrate in both the cases. For mechanically exploited MoS$_2$, the intensity of the Raman peak of the Si substrate is significantly lower as compared with the Raman peak intensity of MoS$_2$ whereas the intensity of Si peak is considerably high in case of ETVA MoS$_2$ film. This indicates that that laser for ETVA MoS$_2$ sample the scattering cross section is lesser than mechanically exfoliated sample which further confirms the presence of oriented stripe like grains in ETVA MoS$_2$ film.

Now a discussion on the line shape asymmetry of $A_g$ mode is in order. This mode was chosen because the peak shape is well defined. In case of bulk MoS$_2$, the full width at half maximum (FWHM) for A$_g$ mode is ~ 3 cm$^{-1}$ whereas its value is about three times higher for the ETVA MoS$_2$ sample. In bulk materials, the zone centre optical phonon ($q$=0) only contributes to the phonon line shape in Raman spectra. However, as the size of the material decreases significantly in the range of sub nanometer scale, the zone boundary phonons also start to contribute in line shape and selection rule ($q$=0) relaxes in such cases which leads to red shift and asymmetric broadening. The phonon confinement model, initially proposed by Richeter *et al.* which was latter modified by Campbell and Fauchet for nanomaterials of different geometries, explains such size dependent Raman spectra.[37, 38] This might be useful to explain the Raman line broadening appeared in vertically oriented stripe like grains of MoS$_2$ films.

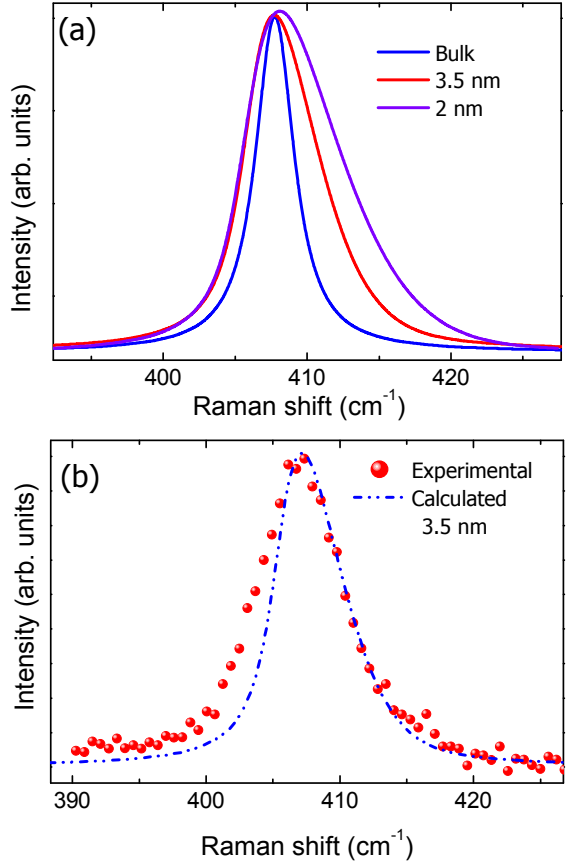

**Fig. 6. (a) Calculated phonon line shape for the $A_{1g}$ mode for two nanoparticle sizes using the phonon confinement model. (b) Comparison of the experimental Raman spectrum and the calculated phonon line shape with particle size 3.5 nm.**

This model considers the contribution of phonons away from the zone center by integrating over the entire Brillouin zone to get the Raman line shape and the phonon amplitude is taken as an exponentially decay function. For a given nanoparticle size *d,* Gaussian confinement function is used which is given by:

$$W(r) = exp\left(\frac{-\alpha r^2}{d^2}\right) \qquad (1)$$

where the magnitude of α determines how fast the wave function decays as one approaches the boundary. In this calculation, $\alpha = 8\pi^2$. The Fourier transformation of the confinement function is the weight factor that estimates the contribution of phonons other than the zone center. The weight factor in this case is

$$|C(q)|^2 = exp\left(\frac{-q^2 d^2}{2\alpha}\right) \qquad (2)$$

The Raman line intensity was calculated by integrating these contributions over the complete Brillouin zone which can be given by

$$I(\omega) = \int \frac{|C(q)|^2}{[\omega - \omega(q)]^2 + \left(\frac{\Gamma_0}{2}\right)^2} d^3q \qquad (3)$$

where $\Gamma_0$ and $\omega(q)$ are the natural line width and the phonon dispersion curve of zone center optical phonon in bulk $MoS_2$. Figure 6 (a) shows the calculated phonon lineshape of $A_g$ mode for two nanoparticle sizes using phonon confinement model. It may be pointed out that as the particle size decreases the calculated Raman lineshape broadens significantly. Observed line shape is fitted best with 3.5 nm grain size (Fig.6 b) which is in very good accordance TEM results (the average grain size was measured to be less than 5 nm). However, the fitted line shape does not fit well in the low frequency side of the experimental spectrum; the low frequency side of the experimental spectrum is border.

This section discusses on the disagreement of the experimental line shape from phonon confinement model. We first consider the microscopic factors affecting the phonon line shape in bulk material and given by Matthiessen's rule: $\frac{1}{\tau} = \frac{1}{\tau_u} + \frac{1}{\tau_b} + \frac{1}{\tau_m} + \frac{1}{\tau_{e-ph}}$ where the terms on the right side denote inverse of the life time of different microscopic processes and contribute to phonon line shape via Umklapp process, boundary, mass difference and electron phonon scattering, respectively.[39] This is due to the fact that the phonon confinement model does not consider these factors in to account. In the present work, Umpklapp scattering is not significant as the experiments were performed at room temperature. Similarly, the presence of isotopes in our sample can be neglected and hence mass difference contributions to the Raman lines shape. However, isotope effect on the Raman line shape has been observed in graphene. The other two remaining scattering processes can significantly affect the Raman line shape especially in nanocrystals. Apart from the conventional boundary scattering, the defects in nanocrystals can act as scattering centers and contribute to the Raman line shape.

Recently, we have demonstrated the temperature dependent Raman studies to understand the lattice anharmonicity in high quality large scale few layer MoS$_2$ thin film. It is also appropriate to investigate the lattice anharmonicity in ETVA MoS$_2$ film. Thus we performed temperature dependent Raman studies in the temperature range 83 to 423K. Higher temperatures were not considered as it could damage it. Figure 7 (a) shows the $A_g$ and $E_{2g}^1$ modes in this temperature range. A blue shift in the peak position of the $A_{1g}$ mode was observed as the temperature was decreased.

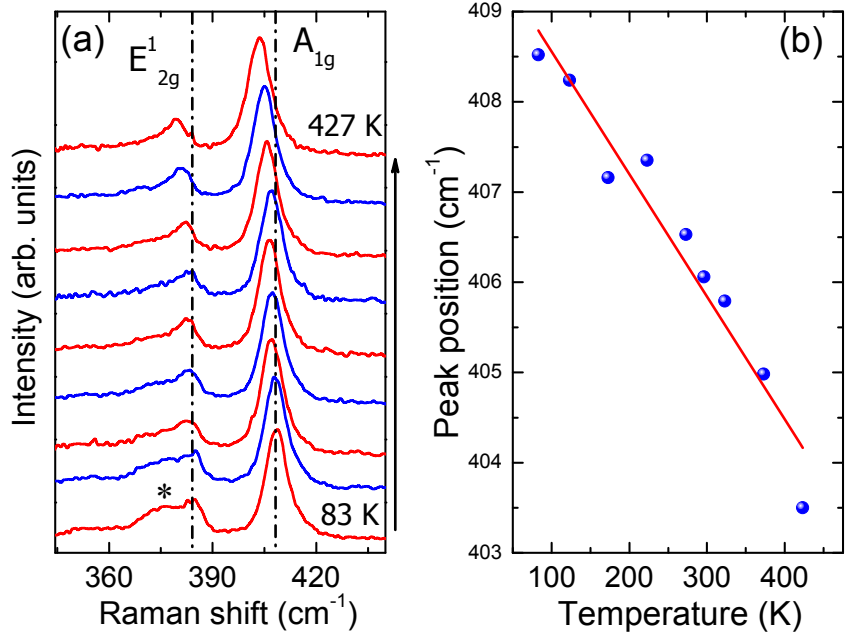

**Fig. 7. (a) Temperature dependent Raman spectra for ETVA MoS$_2$ thin film. (b) Plot of the temperature versus the A$_{1g}$ peak position.**

Surprisingly, a new peak (marked as * in the figure) at about 373 cm$^{-1}$ was observed at the low energy side of E$^1_{2g}$ mode. This was not observed for high quality few layer MoS$_2$. Sekine et al. has observed a similar trend for bulk MoS$_2$ by tuning the laser excitation energy across the A and B excitonic levels.[40] The new peak associated with the E$^1_{2g}$ mode at low temperature was assigned to the E$^2_{1u}$ mode which is a Raman inactive mode. This peak appears due to resonance effect and the small frequency splitting of the Davydov pair (E$^2_{1u}$, E$^1_{2g}$) is caused by very weak interlayer interaction. Figure 7 (b) shows the plot of the A$_{1g}$ mode peak position as a function of temperature. The data was fitted using the linear equation $\omega(T) = \omega_o + \chi T$ where $\omega_o$ is vibrational frequency at absolute zero temperature and $\chi$ is first order temperature coefficient of $A_{1g}$ mode. The fitted numerical value of $\chi$ (-1.36x10$^{-2}$) matches well with the bulk MoS$_2$.

In conclusion, nanocrystalline thin films of $MoS_2$, comprised of a mixture of edge terminated vertically aligned (ETVA) and (001)-oriented regions, were synthesized on large insulating substrates. Weak and broad excitonic peaks, attributed to defects associated with nanocrystals, were observed both in the absorption and transmission spectra due to the spin-orbit splitting of the valence band. The theoretically calculated band structure revealed that with a decrease in the number of layers, the band gap of $MoS_2$ increased from the room temperature Raman spectroscopic studies, it was observed that $E^1_{2g}$ and $A_{1g}$ peaks in ETVA $MoS_2$ are significantly larger than those of bulk $MoS_2$. The Raman line broadening of the $A_{1g}$ mode was analyzed using the phonon confinement model, and the temperature coefficient of the $A_{1g}$ mode was found to be $-1.36 \times 10^{-2}$. The calculated particle size was found to be in good agreement with TEM observations.

**Acknowledgements:** The authors acknowledge financial support from DOE (grant DE-FG02-ER46526). Anand. P. S. Gaur thanks NSF fellowship (grant NSF-RII-1002410).